\documentclass[manuscript]{aastex}

\shorttitle{Anisotropic Transport Effects}
\shortauthors{Devlen}

\begin{document}


\title{THE ANISOTROPIC TRANSPORT EFFECTS ON THE DILUTE PLASMAS}


\author{EBRU DEVLEN}
\affil{University of Ege, Faculty of Science, Department of Astronomy \& Space Sciences,\\
       \noindent\hspace*{32pt}Bornova, 35100, Izmir, TURKEY}

\email{ebru.devlen@ege.edu.tr}




\begin{abstract}

We examine the linear stability analysis of a hot, dilute and differentially rotating plasma by considering anisotropic transport effects. In the dilute plasmas, the ion Larmor radius is small compared with  its collisional mean free path. In this case, the transport of heat and momentum along the magnetic field lines  become important. This paper presents a novel linear instability that may more powerful and greater than ideal magnetothermal instability (MTI) and ideal magnetorotational instability (MRI) in the dilute astrophysical plasmas. This type of plasma is believed to be found in the intracluster medium of galaxy clusters and radiatively ineffective accretion flows around black holes. We  derive the dispersion relation of this instability and obtain the instability condition. There is at least one unstable mode that is independent of  the temperature gradient direction for a helical magnetic field geometry. This novel instability is driven by the gyroviscosity coupled with differential rotation. Therefore we call it as gyroviscous modified magnetorotational instability (GvMRI). We examine how the instability depends on signs of the temperature gradient and  the gyroviscosity, and  also on the magnitude of the thermal frequency and on the values of the pitch angle. We provide a detailed physical interpretation of obtained results. The GvMRI is applicable  not only to the accretion flows and intracluster medium but also to the transition region between cool dense gas and the hot low-density plasma in stellar coronae, accretion disks, and the multiphase interstellar medium because of being independent of the  temperature gradient direction.

\end{abstract}





\section{INTRODUCTION}

In recent years, MRI has been recognized as a powerful source of angular momentum transport in accretion discs. An accretion disc with angular
velocity decreasing outward and threaded by a weak magnetic
field is linearly unstable  (Balbus \& Hawley 1991). Local and global simulations of Keplerian discs showed that the MRI leads to a turbulence and therefore  energy and angular momentum is transported outward (Balbus 2003). Similiarly, the dilute, stratified plasma is bouyantly unstable when the temperature  increases in the direction of gravity. Heat is transported mainly along magnetic field lines in  such a medium (Balbus 2000, 2001). The MTI has been studied with nonlinear simulations. Parrish and Stone (2005) investigated the nonlinear evolution of the MTI and showed that the instability causes turbulence and heat transport. They noted that
MTI may  explain the almost isothermal temperature profile observed in the outer part of  X-ray emitting regions in intracluster medium (ICM) of galaxy clusters  and the structure of radiatively inefficient accretion flows. 
In the MRI and MTI, the weak magnetic fields act as to turn free-energy gradients into sources of the instability. 

If the plasma is sufficiently dilute, the viscous stress tensor  is also anisotropic (Braginskii 1965). Balbus (2004) showed that the viscous stress tensor can cause a strong instability in the dilute astrophysical discs. He showed that the maximum growth rate of instability exceeds that of the MRI.  But, he only took parallel  component of the stress tensor into account. The cause of the magnetoviscous instability is the same with MTI:  Initially magnetic field lines are isorotational (isothermal  in the case of MTI). Perturbed magnetic field lines are stretched out in the direction of the angular velocity (temperature) gradient. Thus, angular momentum (heat) is transferred from one fluid element to another. One fluid element at smaller radius drops down to smaller radii and other element moves to more distant radii. So, the field lines become more bent  and the process runs away (Islam \& Balbus 2005).

Indeed, to understand the true nature of the  dilute astrophysical plasmas, we consider  magnetohydrodynamic (MHD) equations which  include the terms describing  transport of heat and momentum by thermal conduction and viscosity (Braginskii 1965). Dilute means that the ion Larmor radius ($r_{Li}$)  is small compared with a mean free path  ($\lambda_{i}$) and any macroscopic length scales in the plasma. This is tantamount to saying that the ion cyclotron frequency ($\omega_{ci}$) is much greater than the ion-ion  collision frequency ($\nu_{i}$ ).  Under this condition,   parallel heat conduction of electrons  is much larger than that of ions by the factor  $(m_{i}/m_{e})^{1/2}$ and the parallel viscosity of ions is much larger than that of electrons by the same factor. Therefore, an anisotropic electron heat conduction  and an anisotropic ion viscosity must be taken into account in the MHD  equations. Ramos (2003) obtained dynamic evolution equations of the parallel heat fluxes in a collisionless magnetized plasma. Besides, he noted that neglecting the parallel heat fluxes in low collisional regime cannot be justified physically. He emphasized  that one must consider the contribution of  the gyroviscosity in the stress tensor for consistency in the analysis.  Ramos (2005) presented  the fluid moment equations with Finite Larmor Radius (FLR) effect for collisionless magnetized plasmas. His analysis included  the gyroviscous stress, the pressure anisotropy and  the anisotropic heat fluxes. He claimed that his formalism can be applicable for arbitrary magnetic field geometry and arbitrary plasma pressure and also fully electromagnetic nonlinear dynamics. He extended previous study considering collisional terms based on full Fokker-Planck operators for non-Maxwellian distribution functions. The  low collisional  regime of interest is described with two small parameters: the ratio of the electron to the ion masses is comparable to $\delta$ (i.e., $(m_{e}/m_{i})^{1/2}\la\delta\ll 1$)   and the ratio of the ion collision to cyclotron frequencies is smaller than $\delta^2$ (i.e., $\nu_{i}/\omega_{ci}\la \delta^2$),  where $\delta$   is the fundamental expansion parameter which is the ratio of the ion Larmor radius to the shortest macroscopic length scale (Ramos 2007).

Recently, Ferraro (2007) examined the FLR  effects on the magnetorotational instability. He showed that the FLR effects are dominant compared to the other effects which includes the Hall effect  in the limit of weak magnetic fields.   He restricted his analysis in a vertical magnetic field geometry. He found that the growth rate of unstable mode is around $\Omega$ when the ratio of the gyroviscous force to the magnetic tension force is greater than zero. But,  when the ratio of the gyroviscous force to the magnetic tension force is smaller than zero, there is no unstable mode. Devlen \& Pek\"{u}nl\"{u} (2010, Paper I) investigated the stability properties of  weakly magnetized, dilute plasmas by considering combined effects of gyroviscosity and parallel viscosity which are components of the stress tensor in the presence of a helical magnetic field geometry. They showed that although the parallel viscosity is greater than gyroviscosity under the condition of dilute plasma, it hasn't any effect on the instability condition and growth rates. Also they showed that the powerful  instability emerges due to finite Larmor Radius effects. They estimated that the growth rates of this GvMRI varied in the range of  $0.5\Omega-3\Omega$  for the different values of pitch angles which is the angle between the magnetic field vector and the $\phi$ axis of the coordinate system. When the ratio of the gyroviscous force to the magnetic tension force is smaller than zero (i.e., in the case of  $\mathbf{\Omega}\uparrow\downarrow\mathbf{B}_{z}$), they found that there are unstable modes with growth rates around $2\Omega$. This result was contrary to that of Ferraro's (2007).

To clearly comprehend the  dynamics of dilute astrophysical plasmas, all the anisotropic transport effects must be taken into account. Therefore, in this work, we extend our previous study (Paper I) by considering anisotropic electron heat conduction term.  We show that weakly magnetized, differentially rotating dilute plasmas are unstable in the presence of the parallel viscosity, gyroviscosity and thermal conduction. This novel instability is extremely powerful and occurs at  all wavenumbers.

This paper is organized as follows. In the next section (\S2), 
we give the linearization of the MHD equations used in our analysis. We examine the physical structure of modes. We derive the dispersion relation of instability and the instability criterion.  In \S3, we examine the numerical solutions of the dimensionless dispersion relation and finally in \S4,  we discuss the physical interpretation of instability and summarize the  our results .

\section{LINEAR STABILITY ANALYSIS}

\subsection{Dilute Plasma Properties}

The plasma fulfilling the conditions $r_{Li}\ll\lambda_{i}$ and $\epsilon\equiv\omega_{ci}\tau_{i}\gg 1$ is called  dilute plasma, where  $r_{Li}$ is the ion Larmor radius, $\lambda_{i}$ is 
the ion collision mean free path, $\omega_{ci}$ is the ion cyclotron frequency and $\tau_{i}=1/\nu_{i}$ is the inverse of the ion-ion collision frequency. The presence of the magnetic field introduces anisotropy to the medium.  Cyclotron  frequencies of the plasma species and the velocity gradients at macroscopic scales are the sources of anisotropy. 

If plasma consists only of hydrogens then   $\epsilon$ may be taken as (Spitzer 1962)

\begin{equation}
\epsilon=\left({{1.09\times10^5}\over{n}}\right) {T_4^{3/2}B_{\mu G}\over\ln\Lambda},
\end{equation} 

where $n$ is the proton number density in cm$^{-3}$, $T_4$ the temperature in units of $10^4$ K, $B_{\mu G}$ is the magnetic field in units of $10^{-6}$ and $\ln\Lambda$ is the Coulomb logarithm.  The condition of $\epsilon\gg1$ is fulfilled even in the presence of a  very weak field with $n\la1$ and $T_4\ga1$. 

Under these conditions, the plasma dynamics described by MHD equations should include the anisotropic terms accounting for the free flow of particles along the magnetic field lines (Braginskii 1965). Ion parallel viscosity is higher by a factor $(m_{i}/m_{e})^{1/2}$ than that of electrons. So, the viscosity of the dilute plasma is determined mainly by the ions. Even if the ion viscosity is very small, in a rotating system, it may  become very important (Balbus 2004). Similarly, since the electron contribution to the heat flux is higher than the ion contribution by a factor of $(m_{i}/m_{e})^{1/2}$, the ion contribution may be considered as negligible. Since the electrons have mean free paths much longer than their gyro-radii in the dilute plasma, the thermal conductivity is strongly anisotropic. That is, in the astrophysical dilute plasma threaded even by a weak magnetic field,  the momentum by ions and heat flux by electrons is transported primarily along the magnetic field lines.

\subsection{Basic Equations}

In an attempt to investigate parallel viscosity, gyroviscosity and heat flux in a dilute plasma, one should consider the two-fluid equations. Below are the standart extended MHD equations which are  obtained by using two-fluid equations including stress tensor, $\mathbf{\Pi}$ , and  the heat flux $\mathbf{Q}$ (see Appendix A):

\begin{equation}
{d\rho\over dt}+\rho\nabla\cdot\mathbf{v}=0,
\end{equation}

\begin{equation}
\rho{d{\mathbf{v}}\over dt}=-\nabla P-\nabla\cdot\mathbf{\Pi}+{(\nabla\times\mathbf{B})\times\mathbf{B}\over c}+\rho\mathbf{g},
\end{equation}

\begin{equation}
{\partial{\mathbf{B}}\over\partial{t}}=\nabla\times \left(\mathbf{v}\times\mathbf{B}\right),
\end{equation}

\begin{equation}
{dP\over dt}+{5\over 3}P\left(\nabla\cdot\mathbf{v}\right)=-{2\over 3}\nabla\cdot\mathbf{Q},
\end{equation}

where $\rho$ is the mass density,  $\mathbf{v}$ is the fluid velocity,  $P$ is the scalar pressure,   $\mathbf{\Pi}$  is the stress tensor, $\mathbf{B}$ is the magnetic field, $\mathbf{g}$  is the gravitational acceleration, $\mathbf{Q}$ is the heat flux, and   $d/dt=\partial/\partial t+\mathbf{v}\cdot\nabla$ is a Lagrangian derivative.

Stress tensor have three  components which are called as parallel $(\parallel)$, perpendicular $(\perp)$  and the gyroviscous $(gv)$ (Braginskii 1965). Perpendicular  viscosity is greater than the parallel viscosity by a factor of $(r_{Li}/l)^{2}$, therefore it may not be taken into account; where $r_{Li}$ is the Larmor radius and $l$ is the mean free path of the particles in a dilute plasma. So, we used parallel and gyroviscous components of the stress tensor which are given by

\begin{equation}
\mathbf{\Pi}^{v}=0.96{P_{i}\over 2\nu_{i}}\left(\mathbf{I}-3\mathbf{\hat{b}}\mathbf{\hat{b}}\right)\left(\mathbf{\hat{b}}\cdot\mathbf{W}\cdot\mathbf{\hat{b}}\right),
\end{equation}

\begin{equation}
\mathbf{\Pi}^{gv}={P_{i}\over 4\omega_{ci}}\left[\mathbf{\hat{b}}\times\mathbf{W}\cdot\left(\mathbf{I}+3\mathbf{\hat{b}}\mathbf{\hat{b}}\right)+\left[\mathbf{\hat{b}}\times\mathbf{W}\cdot\left(\mathbf{I}+3\mathbf{\hat{b}}\mathbf{\hat{b}}\right)\right]^{T}\right],
\end{equation}

where  $ \mathbf{\hat{b}}=\mathbf{B}/B$, $ \mathbf{\omega_{ci}}=eB/m_{i}c$  are the unit vector along the magnetic field and the cyclotron frequency. $ \nu_{i}$ is the ion collision frequency. $\mathbf{W}=\nabla\mathbf{v}+(\nabla\mathbf{v})^{T}-2/3\mathbf{I}\left(\nabla\cdot\mathbf{v}\right)$ is the rate of strain tensor.

In a dilute astrophysical plasma heat flux $\mathbf{Q}$ is dominantly along the magnetic field lines. The parallel heat flux is given by

\begin{equation}
 \mathbf{Q}=-\chi_{C}\mathbf{\hat{b}}\left(\mathbf{\hat{b}}\cdot\nabla \right)T, 
\end{equation}

where  $\chi_{C}$ is the Coulomb conductivity given by Spitzer (1962) as $\chi_{C}\simeq6\times10^{-7}T^{5/2}$ $ergs cm^{-1} K^{-1}$.

\subsection{Linearized Expressions for Perturbed Quantities}

We apply a standart Wentzel-Kramers-Brillouin (WKB) perturbation analysis on the equilibruim state. To this analysis,  all the variables in the MHD equations are denoted by sums of an equilibrium value (denoted with a ``0'' subscript) and a small perturbed quantity (denoted with $\delta$)

\begin{eqnarray}
\rho=\rho_{0}+\delta\rho, \nonumber\\ \mathbf{v}=\mathbf{v}_0+\delta\mathbf{v}, \nonumber\\
P=P_{0}+\delta P,\nonumber\\
\mathbf{\Pi}=\mathbf{\Pi}_0+\delta\mathbf{\Pi}, \nonumber\\
\mathbf{B}=\mathbf{B}_0+\delta\mathbf{B},\nonumber\\
\mathbf{Q}=\mathbf{Q}_0+\delta\mathbf{Q}. 
\end{eqnarray}
 
Substituting the formulae in equation (9) into equations (2)-(8) and retaining only terms up to linear order in perturbations,  the linearized perturbation equations are obtained as

\begin{equation}
\nabla\cdot\delta v=0,
\end{equation}

\begin{equation}
{\partial\delta{\mathbf{v}}\over \partial t}+\delta {\mathbf{v}}\cdot\nabla{\mathbf{v}}={\delta\rho\over\rho^2}\nabla P-\nabla\cdot\mathbf{\delta\Pi}-{1\over\rho}\nabla\left(\delta P+{\delta\mathbf{B}\cdot\mathbf{B}\over 4\pi}\right)+{(\mathbf{B}\cdot\nabla)\over 4\pi\rho}\delta\mathbf{B},
\end{equation}

\begin{equation}
{\partial{\delta\mathbf{B}}\over\partial{t}}=\nabla\times \left(\delta\mathbf{v}\times\mathbf{B}\right)+\nabla\times \left(\mathbf{v}\times\delta\mathbf{B}\right),
\end{equation}

\begin{equation}
{5\over 3}{\partial\over\partial t}{\delta\rho\over\rho}-\delta\mathbf{v}\cdot\nabla\ln P\rho^{-5/3}={2\over 3P}\nabla\cdot\delta\mathbf{Q}.
\end{equation}

The perturbed parallel and gyroviscous components of the stress tensor are

\begin{eqnarray}
\delta\mathbf{\Pi}^{v}=0.96{P_{i}\over 2\nu_{i}}\left[\begin{array}{l}
(\mathbf{I}-3\mathbf{\hat{b}}\mathbf{\hat{b}})(\delta\mathbf{\hat{b}}\cdot\mathbf{W}\cdot\mathbf{\hat{b}})+(\mathbf{I}-3\mathbf{\hat{b}}\mathbf{\hat{b}})(\mathbf{\hat{b}}\cdot\delta\mathbf{W}\cdot\mathbf{\hat{b}})\\
+(\mathbf{I}-3\mathbf{\hat{b}}\mathbf{\hat{b}})(\mathbf{\hat{b}}\cdot\mathbf{W}\cdot\delta\mathbf{\hat{b}})
\end{array}
\right],
\end{eqnarray}

\begin{eqnarray}
\delta\mathbf{\Pi}^{gv}={P_{i}\over 4\omega_{ci}}\left\{\begin{array}{l}
\left[\begin{array}{l}
\delta\mathbf{\hat{b}}\times\mathbf{W}\cdot\left(\mathbf{I}+3\mathbf{\hat{b}}\mathbf{\hat{b}}\right)+\mathbf{\hat{b}}\times\delta\mathbf{W}\cdot\left(\mathbf{I}+3\mathbf{\hat{b}}\mathbf{\hat{b}}\right)\\
+\mathbf{\hat{b}}\times\mathbf{W}\cdot3\delta\mathbf{\hat{b}}\mathbf{\hat{b}}+\mathbf{\hat{b}}\times\mathbf{W}\cdot3\mathbf{\hat{b}}\delta\mathbf{\hat{b}}
\end{array}
\right]\\
\left[\begin{array}{l}
\delta\mathbf{\hat{b}}\times\mathbf{W}\cdot\left(\mathbf{I}+3\mathbf{\hat{b}}\mathbf{\hat{b}}\right)+\mathbf{\hat{b}}\times\delta\mathbf{W}\cdot\left(\mathbf{I}+3\mathbf{\hat{b}}\mathbf{\hat{b}}\right)\\
+\mathbf{\hat{b}}\times\mathbf{W}\cdot3\delta\mathbf{\hat{b}}\mathbf{\hat{b}}+\mathbf{\hat{b}}\times\mathbf{W}\cdot3\mathbf{\hat{b}}\delta\mathbf{\hat{b}}
\end{array}
\right]^T
\end{array}
\right\}.
\end{eqnarray}

The perturbed heat flux is given by

\begin{equation}
\delta\mathbf{Q}=-\chi_{C}\left[\mathbf{\hat{b}}\left(\delta\mathbf{\hat{b}}\cdot\nabla \right)T-i\mathbf{\hat{b}}\left(\mathbf{\hat{b}}\cdot k \right)\delta T\right].
\end{equation}

The perturbed unit vector of the magnetic field is given by $\delta\mathbf{\hat{b}}=\delta(\mathbf{B}/B)=\delta\mathbf{B}/B-\mathbf{\hat{b}}(\delta B/B)$.

\subsection{The Physical Structure of Modes}

Before we obtain the dispersion relation which includes all anisotropic transport effects  for general axisymmetric disturbances, let us take a glance at the physical meaning of finite Larmor radius effect and modes which emerge in the plasma.  We consider  the local stability of a  uniformly rotating dilute plasma included  only a vertical magnetic field,  $\mathbf{B}=B\mathbf{\hat z}$.  We ignore parallel viscosity, heat flux and radial stratification.  We restrict ourself to plane wave perturbations that depend only on $z$, i.e., of the form $\exp(ikz-i\omega t)$. Thus, the radial and  azimuthal components of the linearized motion equation are obtained as 

\begin{equation}
-i\omega\delta v_R -(2\Omega-4V_{gyro})\delta v_{\phi}-{ikB\over{4\pi\rho}}\delta B_{R}=0,
\end{equation}

\begin{equation}
-i\omega\delta v_\phi +(2\Omega-4V_{gyro})\delta v_{R}-{ikB\over{4\pi\rho}}\delta B_{\phi}=0,
\end{equation}

where $V_{gyro}=k_{z}^2P/4\omega_{ci}\rho$ is the inverse time scale of the gyroviscous stress. The same components of the linearized magnetic induction equation are

\begin{equation}
-i\omega\delta B_R -ikB\delta v_{R}=0,
\end{equation}

\begin{equation}
-i\omega\delta B_\phi -ikB\delta v_{\phi}=0.
\end{equation}

The gyroviscous force  introduces a term   like a Coriolis term in the equation of motion. The dispersion relation is 

\begin{equation}
\omega^4-\omega^2\left[2k^2v_A^2+4\Omega^2\left(1-{k^2v_D^2\over\Omega^2}\right)^2\right]+k^4v_A^4=0,
\end{equation}

where $v_A^2=B^2/4\pi\rho$ is the Alfven velocity, $v_D^2=P\Omega/2\omega_{ci}\rho$ is the drift velocity.

FLR stress results from changes in particle drift velocities across a gyro-orbit. This stress  gives rise to distortions of particle orbits  and guiding-center drift. Kaufman (1960) presented a detailed discussion of this stress. Because ions and electrons have different Larmor radii, they move differently due to FLR effects. This  different motion  gives rise to charge separation. Then it produces a finite parallel electric field.   Physically, the FLR effects introduce a drift wave that convects the perturbations along the velocity gradient. 

If the angular velocity and magnetic field  vectors are oriented in the same sense, the term including $V_{gyro}$ is positive. From equations (17)-(20), one   can easily see that  the induced  drift motion is opposite to  the Coriolis force. Thus, the dynamical epicycle is slowed and magnetic tension force is effectively  increased. This, in turn, increases the angular momentum transfer. The result is an instability. If the angular velocity and magnetic field  vectors are counter aligned,  the signs of these effects should reverse.

In the limit $\Omega\rightarrow 0$, the dispersion relation is obtained as

\begin{equation}
\omega^4-\omega^2\left[2k^2v_A^2+k^4\left({Pmc\over eB\rho}\right)^2\right]+k^4v_A^4=0.
\end{equation}

The solutions of equation (22) give two roots:  

\begin{equation}
\omega^2={1\over 2}\left[2k^2v_A^2+k^4\left({Pmc\over eB\rho}\right)^2\right]\pm{1\over 2}\left[k^4\left({Pmc\over eB\rho}\right)^2\left(4k^2v_A^2+k^4\left({Pmc\over eB\rho}\right)^2\right)\right]^{1/2}.
\end{equation}

One of the roots  describes pure drift mode at large wavenumbers.  For small wavenumbers (low frequencies) the other  root is 

\begin{equation}
\omega^2=k^2v_A^2\left(1\pm k^2{{Pmc / eB\rho}\over kv_A}\right),
\end{equation}

corresponding to Alfven waves with the gyroviscous force producing a small frequency-splitting of the Alfven wave (see Figure 1).

\begin{figure}
\epsscale{.50}
\plotone{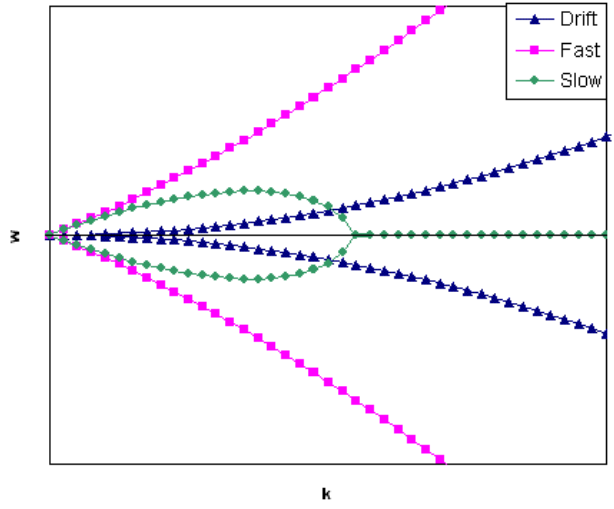}
\caption{A wavenumber-frequency diagram in the absence of rotation.  Forward and backward waves correspond to the positive and negative values, respectively.}
\end{figure}

\begin{figure}
\epsscale{.50}
\plotone{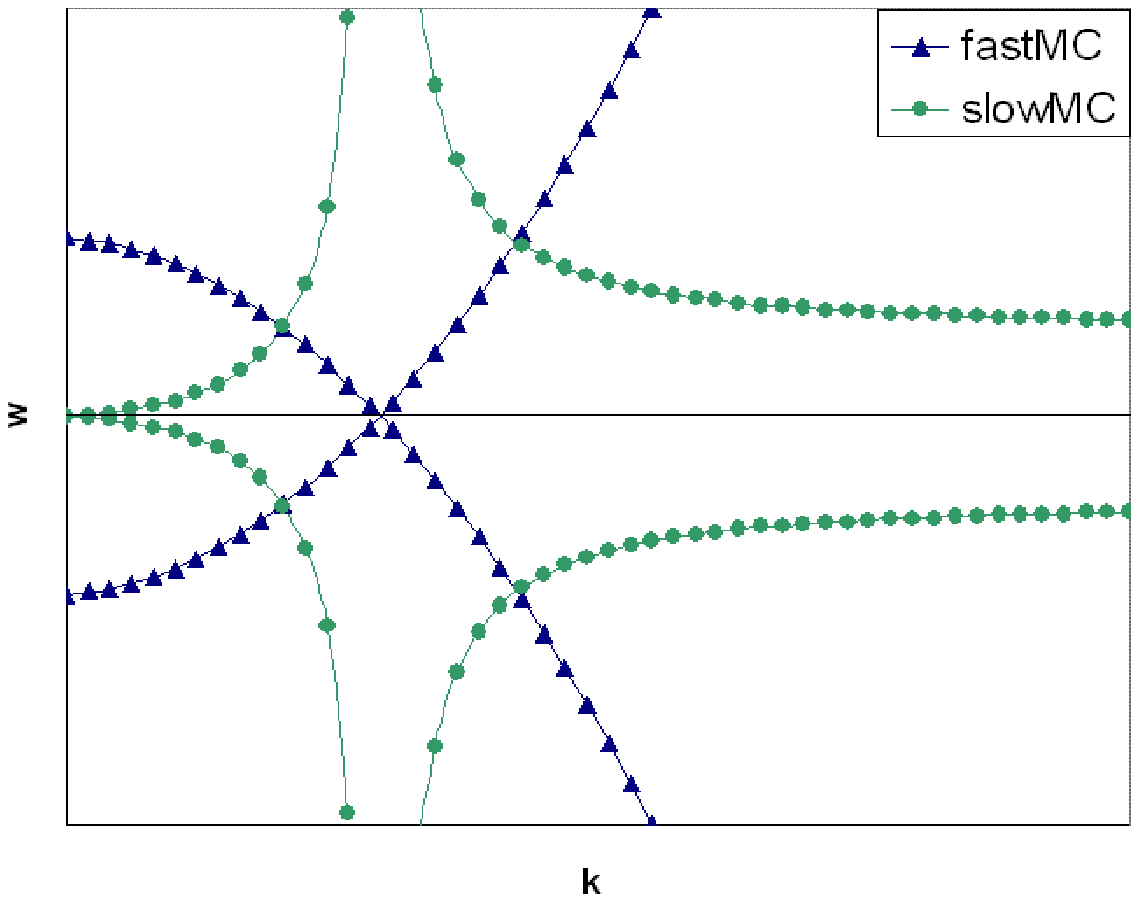}
\caption{A wavenumber-frequency diagram in the presence of uniform rotation.  Forward and backward waves correspond to the positive and negative values, respectively.}
\end{figure}

One may rewrite the dispersion relation (21) in the presence of the uniform rotation:

\begin{equation}
\omega^2\pm \omega\omega_I\left(1-{4\omega_D^2\over\omega_I^2}\right)-\omega_A^2=0,
\end{equation}

where drift wave frequency is $\omega_D^2=Pmc(\mathbf{k}\cdot\mathbf{\Omega})^2/2e\rho(\mathbf{\Omega}\cdot\mathbf{B})$, Alfven wave frequency is  $\omega_A^2=(\mathbf{k}\cdot\mathbf{B}/4\pi\rho)$,  pure inertial wave frequency is $\omega_I=2(\mathbf{k}\cdot\mathbf{\Omega})/k$ for general magnetic field geometries and wavenumbers (Mofatt 1978).  
 The solutions of the Equation (25) are
 
\begin{equation}
\omega=\pm{1\over 2}\omega_I\left(1-{4\omega_D^2\over\omega_I^2}\right)\pm{1\over 2}\left[\omega_I^2\left(1-{4\omega_D^2\over\omega_I^2}\right)^2+4\omega_A^2\right]^{1/2}.
\end{equation}

For $(\mathbf{k}\cdot\mathbf{\Omega})=0$, it follows $\omega_+=\omega_-=\omega_A$. If the first term is greater than the second one in the brackets in the equation (26), then it is possible to carry out a Taylor series expansion of this equation. One then finds a very clear splitting of the fast and slow wave frequencies, 

\begin{equation}
\omega_+=\pm\omega_I\left(1-{4\omega_D^2\over\omega_I^2}\right)\left(1+{\omega_A^2\over 
{\omega_I^2\left(1-{4\omega_D^2\over\omega_I^2}\right)^2}}\right),
\end{equation}

and

\begin{equation}
\omega_-=\pm{\omega_A^2\over 
{\omega_I\left(1-{4\omega_D^2\over\omega_I^2}\right)}}.
\end{equation}

\begin{figure}
\epsscale{.5}
\plotone{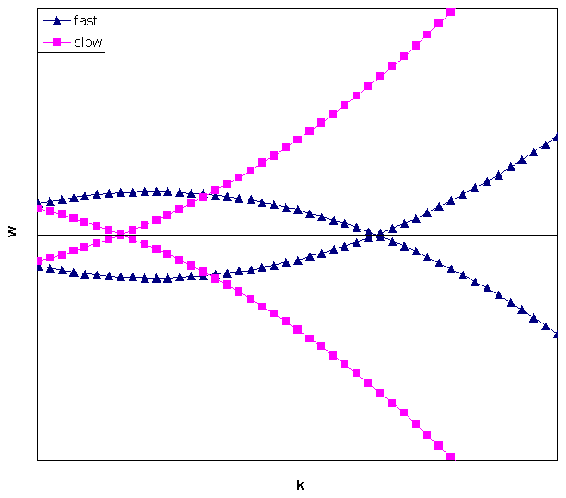}
\caption{A wavenumber-frequency diagram in the presence of uniform rotation.   Forward and backward waves correspond to the positive and negative values, respectively.}
\end{figure}

respectively. These waves result as a combination of the inertial and Alfven waves. They are referred to as magnetocoriolis (MC)  waves. When the magnetic field and rotation axis are aligned, the MC wave which imparts a circularly polarized component to the velocity perturbation  is  in the anticlockwise direction. Therefore Coriolis and Lorentz forces are in phase.  Resulting force causes  inertial acceleration. This mode known as fast MC wave (i.e., $\omega_+$ mode). Indeed, in the equations (27) and (28) the inertial wave is coupled with the drift mode.  So, this mode has a larger frequency compared that of the  pure fast MC wave. For  $\omega_-$  mode,  the MC wave which  imparts a circularly polarized component to the velocity perturbation  is  in the clockwise direction. Hence, Coriolis and Lorentz forces are out of  phase. Resulting force is weakened.  This slow MC mode is sometimes referred to as magnetostrophic wave or hydromagnetic-inertial wave (see Figure 2). These waves are especially important in the dynamo problem (Moffatt 1978, Achenson \& Hide 1973). 

When the first term is smaller than the second one in the brackets in equation (26), one obtains modified Alfven waves with frequency splitting after Taylor series expansion: 

\begin{equation}
\omega=\pm\omega_A\left(1\pm{\omega_I\over {2\omega_A}}\left(1-{4\omega_D^2\over\omega_I^2}\right)\right).
\end{equation}

A wavenumber-frequency diagram of these waves is given in Figure 3. In general the angular velocity and magnetic field  vectors will not be parallel and the situation will be more complex. But the physical picture remains the same.

\subsection{Dispersion Relation with All Anisotropic Transport Effects}

We now consider  the axisymmetric behavior of the instability  for helical magnetic field and more general wavenumbers.

\subsubsection{Equilibrium state}

We work in a cylindrical coordinates system, $(R, \phi, z)$.  The plasma is assumed to be thermally stratified in the presence of a uniform gravitational field in the radial direction, $\mathbf{g}=-g\mathbf{\hat{R}}$. The weak magnetic field is taken to be  $\mathbf{B}=(0, B_{0}cos\theta, B_{0}sin\theta)$, where $\theta=\tan^{-1}(B_{z}/B_{\phi})$ which is called ``pitch angle''.  In equilibrium  the field lines are assumed to be isothermal and therefore heat flux is negligible. We consider differentially rotating plasma with a Keplerian velocity profile, i.e.  $v_{\phi}=R\Omega(R)$. For this profile and weak magnetic field, the stress tensor is negligibly small  in the equilibrium for an arbitrary pitch angle (see Appendix B).  Hence,  plasma is in a hydrostatic equilibrium:

\begin{equation}
{\nabla P_{0}\over\rho_{0}}=\mathbf{g}+R\Omega^2.
\end{equation}

\subsubsection{General axisymmetric disturbances}

All the perturbed quantities are assumed to have a space-time dependence  $\exp(i\mathbf{k}\cdot\mathbf{r}+\omega t)$, where  $\mathbf{k}=k_{R}\mathbf{\hat{R}}+k_{z}\mathbf{\hat{z}}$. WKB assumption requires $kR\gg 1$.  The time dependence of the perturbations are assumed as $\exp(\omega t)$. This assumption ensures that all coefficients in the dispersion relation are real.   We work in the Boussinesq limit. In this limit, pressure changes are much smaller than temperature and density changes, i.e.,  $\delta T=-T\left(\delta\rho/\rho\right)$.

The above equations may be written in explicit component form. The perturbation equation of  mass continuity is given by 

\begin{equation}
k_{R}\delta v_{R}+k_{z}\delta v_{z}=0.
\end{equation}

The radial, azimuthal and the axial components of the linearized momentum conservation equation are given by the equations (32), (33) and (34), respectively, 

\begin{eqnarray}
\omega\delta{v}_{R}-2\Omega\delta{v}_{\phi}-{\delta\rho\over\rho^2}{\partial P\over\partial R}+ik_{R}{\delta P\over\rho}+{1\over4\pi\rho}ik_{R}\left({B}_{\phi}\delta{B}_{\phi}+{B}_{z}\delta{B}_{z}\right)
-{1\over4\pi\rho}ik_{z}{B}_{z}\delta{B}_{R}\nonumber\\
-V_{par}\left[\begin{array}{c}{d\Omega\over d\ln R}{1\over\omega}{k_{R}\over k_z}\sin2\theta\delta v_{R}+{k_{R}\over k_z}\sin2\theta \delta v_{\phi}
+2{k_{R}\over k_z}\sin^{2}\theta\delta v_{z}
\end{array}
\right]\nonumber\\
-V_{gyro}\left[\begin{array}{c}
2\sin\theta{d\Omega\over d\ln R}{1\over\omega}({k_{R}^2\over k_{z}^2}-1)\delta v_{R}
+(2\cos\theta {k_{R}^2\over k_{z}^2}+E)\delta v_{z}
-(2{k_{R}^2\over k_{z}^2}\sin\theta+A)\delta v_{\phi}
\end{array}
\right]=0,
\end{eqnarray}

\begin{eqnarray}
\omega\delta{v}_{\phi}+{\kappa^{2}\over 2\Omega}\delta{v}_{R}-{1\over4\pi\rho}ik_{z}\left({B}_{z}\delta{B}_{\phi}\right)
+V_{par}2D\left[{d\Omega\over d\ln R}{1\over\omega}cos\theta\delta v_{R}+cos\theta \delta v_{\phi}+\sin\theta\delta v_{z}\right]\nonumber\\
+V_{gyro}\left[\begin{array}{c}
-2D{d\Omega\over d\ln R}{1\over\omega}
\delta v_{z}
+{d\Omega\over d\ln R}{1\over\omega}B\delta v_{\phi}
+\left[({d\Omega\over d\ln R})^2{1\over\omega^{2}}B
-A+2\sin\theta {k_{R}^2\over k_{z}^2}\right]\delta v_{R}
\end{array}
\right]=0,
\end{eqnarray}

\begin{eqnarray}
\omega\delta{v}_{z}+ik_{z}{\delta P\over\rho}+{1\over4\pi\rho}ik_{z}{B}_{\phi}\delta{B}_{\phi}
+V_{par}\left[F {d\Omega\over d\ln R}{1\over\omega}
\delta v_{R}+F\delta v_{\phi}+C\delta v_{z}\right]\nonumber\\
+V_{gyro}\left[\begin{array}{c}
\left(2{k_{R}^2\over k_{z}^2}\cos\theta+E+4D({d\Omega\over d\ln R})^2{1\over\omega^{2}}\right)\delta v_{R}
+\left(4{k_{R}\over k_{z}}\sin\theta+4D{d\Omega\over d\ln R}{1\over\omega}\right)\delta v_{\phi}\\
+{d\Omega\over d\ln R}{1\over\omega}G\delta v_{z}
\end{array}
\right]=0.
\end{eqnarray}

Similarly, the radial, azimuthal and axial components of the linearized magnetic induction equation are given by the equations (35), (36) and (37) respectively,

\begin{equation}
\omega\delta B_{R}-ik_{z}B_{z}\delta v_{R}=0,
\end{equation}
\begin{equation}
\omega\delta B_{\phi}-ik_{z}B_{z}\delta v_{\phi}-{d\Omega\over d\ln R}\delta B_{R}=0,
\end{equation}
\begin{equation}
\omega\delta B_{z}-ik_{z}B_{z}\delta v_{z}=0.
\end{equation}

Finally the linearized energy equation is,

\begin{equation}
{\delta\rho\over\rho}\left(\omega+V_{ther}\sin^2\theta\right)-\delta v_{R}\left({3\over 5}{d\ln P\rho^{-5/3}\over dR}+V_{ther}{1\over\omega}\sin^2\theta{d\ln T\over dR}\right)=0,
\end{equation}

where $V_{par}=0.96k_{z}^2P/2\nu\rho$ is the inverse time scale of the dissipation due to parallel viscosity, $V_{gyro}=k_{z}^2P/4\omega_{ci}\rho$ is the inverse time scale of the gyroviscous stress and $V_{ther}=2k_{z}^2\chi T/5P$ is the inverse time scale of the dissipation due  to thermal conductivity or thermal frequency. $\kappa^2$ is epicyclic frequency. Other constants which depend on pitch angle $\theta$ are $A=\sin\theta(1-3\cos2\theta)$, $B=\sin\theta(1+9\cos^2\theta-3\sin^2\theta)$, $C=2\sin^2\theta(-1+3\sin^2\theta)$, $D= 3\sin^2\theta\cos\theta$, $E=2\cos\theta(1+3\sin^2\theta)$, $F=2\sin\theta D-\sin2\theta$, $G=\sin\theta(1+3\cos 2\theta)$ and $H=3\sin\theta\cos^2\theta$.

\subsubsection{Dispersion relation}

The set of equations (31)-(38)  are reduced to  form $M\cdot\delta\mathbf{v}=0$, where $M$ is a $3\times3$ matrix. $\mid M\mid=0$ gives nontrivial solution. Thus, the dispersion relation is obtained as

\begin{equation}
{\omega}^{5}+{a}_{4}{\omega}^{4}+{a}_{3}{\omega}^{3}+{a}_{2}{\omega}^{2}+{a}_{1}{\omega}+{a}_{0}=0,
\end{equation}

where 

\begin{equation}
{a}_4=6{V}_{par}s^2{k_{\perp}^2\over k^2}+V_{ther}s^2,
\end{equation}

\begin{eqnarray}
{a}_3=2k_{z}^2v_{A}^{2}+{k_{z}^2\over k^2}{\kappa}^2+{k_{z}^2\over k^2}\tilde{V}_{gyro}^{2}{\Omega}^2(2sy^2-A)^2+\tilde{V}_{gyro}\Omega^{\prime}(G+H)\nonumber\\
+4\Omega^2{k_{z}^2\over k^2}\tilde{V}_{gyro}(2sy^2-A)
+N^{2}{k_{z}^2\over k^2}+6{V}_{par}V_{ther}s^4{k_{\perp}^2\over k^2},
\end{eqnarray}

\begin{eqnarray}
{a}_2=6{V}_{par}s^2k_{z}^2v_{A}^{2}{k_{\perp}^2\over k^2}+{V}_{par}\tilde{V}_{gyro}3s^2G\Omega^{\prime}{k_{\perp}^2\over k^2}
+{k_{z}^2\over k^2}{V}_{par}\Omega^{\prime}2cD+{k_{z}^2\over k^2}\tilde{V}_{gyro}6y\Omega D\Omega^{\prime}
\nonumber\\
+(\tilde{V}_{gyro})^2 3y\Omega D(2sy^2-A)\Omega^{\prime}{k_{z}^2\over k^2}
+{k_{z}^2\over k^2}{V}_{par}N^{2}2cD-{k_{z}^2\over k^2}{V}_{ther}s^{2}{1\over\rho}{\partial P\over\partial R}{\partial\ln T\over\partial R}\nonumber\\
+{V}_{ther}s^{2}\left[\begin{array}{c}
2k_{z}^2v_{A}^{2}+{k_{z}^2\over k^2}{\kappa}^2+{k_{z}^2\over k^2}\tilde{V}_{gyro}^{2}{\Omega}^2(2sy^2-A)^2+\tilde{V}_{gyro}\Omega^{\prime}(G+H)\\
+4\Omega^2{k_{z}^2\over k^2}\tilde{V}_{gyro}(2sy^2-A)
\end{array}
\right],
\end{eqnarray}

\begin{eqnarray}
{a}_1={k_{z}^2\over k^2}y^2\tilde{V}_{gyro}^2 2D^2(\Omega^{\prime})^2+\left[k_{z}^2v_{A}^{2}+\tilde{V}_{gyro}\Omega^{\prime}(G/2+H)\right]\left[k_{z}^2v_{A}^{2}+\tilde{V}_{gyro}\Omega^{\prime}G/2+{k_{z}^2\over k^2}\Omega^{\prime}\right]\nonumber\\
+{k_{z}^2\over k^2} N^{2}\left[k_{z}^2v_{A}^{2}+\tilde{V}_{gyro}\Omega^{\prime}\left({G\over 2}+H\right)\right]-{k_{z}^2\over k^2}{V}_{ther}s^{2}{V}_{par}2cD{1\over\rho}{\partial P\over\partial R}{\partial\ln T\over\partial R}\nonumber\\
+{V}_{ther}s^{2}\left[\begin{array}{c}
6{V}_{par}s^2k_{z}^2v_{A}^{2}{k_{\perp}^2\over k^2}+{V}_{par}\tilde{V}_{gyro}3s^2G\Omega^{\prime}{k_{\perp}^2\over k^2}
+{k_{z}^2\over k^2}{V}_{par}\Omega^{\prime}2cD+{k_{z}^2\over k^2}\tilde{V}_{gyro}6y\Omega D\Omega^{\prime}\\
+(\tilde{V}_{gyro})^2 3y\Omega D(2sy^2-A)\Omega^{\prime}{k_{z}^2\over k^2}\end{array}
\right],
\end{eqnarray}

\begin{eqnarray}
{a}_0={V}_{ther}s^{2}\left[\begin{array}{c}
{k_{z}^2\over k^2}y^2\tilde{V}_{gyro}^2 2D^2(\Omega^{\prime})^2+\left[k_{z}^2v_{A}^{2}+\tilde{V}_{gyro}\Omega^{\prime}(G/2+H)\right]\\
\times\left[k_{z}^2v_{A}^{2}+\tilde{V}_{gyro}\Omega^{\prime}{G\over 2}+{k_{z}^2\over k^2}\Omega^{\prime}-{k_{z}^2\over k^2}{1\over\rho}{\partial P\over\partial R}{\partial\ln T\over\partial R}\right]\end{array}
\right],
\end{eqnarray}

where $y=k_R/k_z$, $\Omega^{\prime}=d\Omega^2/d\ln R$, $s=sin\theta$, $c=cos\theta$, $k_{\perp}^2=k_{R}^2+k_{z}^2\cos^2\theta $ and $\tilde{V}_{gyro}=k_{z}^2P_{i}/4\Omega\omega_{ci}\rho$. $N^2=(3/5\rho)(\partial P/\partial R)(\partial\ln P\rho^{-5/3}/\partial  R)$ is Brunt-V\"{a}is\"{a}l\"{a} frequency. 

The above dispersion relation is reduced to several previously obtained relations in the appropriate limits. Taking  $V_{par}=V_{gyro}=V_{ther}=\theta=N^2=0$ , one recovers the result of Balbus \& Hawley (1991). Taking $V_{gyro}=V_{ther}=\theta=N^2=0$, one recovers the result of Islam \& Balbus (2005). 
Taking $N^2=0$ and $V_{ther}=0 $, then one recovers the dispersion relation given by Paper I (2010). If  one sets $V_{par}=V_{gyro}=\theta=0$, after some algebra, one recovers the dispersion relations given by Balbus (2001).

\subsection{Instability Criterion}

The solution of the Equation (39) gives five modes that exist in weakly magnetized, dilute plasmas. By analysis of the Routh-Hurwitz criterion, the instability criterion is given by

\begin{eqnarray}
a_{0}=\left[k_{z}^2v_{A}^{2}+\tilde{V}_{gyro}\Omega^{\prime}({G\over 2}+H)\right]\left[k_{z}^2v_{A}^{2}+{k_{z}^2\over k^2}\Omega^{\prime}+\tilde{V}_{gyro}\Omega^{\prime}{G\over 2}-{k_{z}^2\over k^2}{1\over\rho}{\partial P\over\partial R}{\partial\ln T\over\partial R}\right]\nonumber\\
+{k_{z}^2\over k^2}y^2\tilde{V}_{gyro}^2 2D^2(\Omega^{\prime})^2<0.
\end{eqnarray}

This complex criterion is simplified,  if one  considers vertical magnetic field. In this case  $\theta=90^{\circ}, D=0, H=0, G=-2$  and the instability criterion is reduced to 

\begin{eqnarray}
\left[k_{z}^2v_{A}^{2}-\tilde{V}_{gyro}{d\Omega^2\over d\ln R}\right]
\left[k_{z}^2v_{A}^{2}+{k_{z}^2\over k^2}{d\Omega^2\over d\ln R}-\tilde{V}_{gyro}{d\Omega^2\over d\ln R}-{k_{z}^2\over k^2}{1\over\rho}{\partial P\over\partial R}{\partial\ln T\over\partial R}\right]<0.
\end{eqnarray}

In the absence of gyroviscosity and heat conduction, this criterion is reduced to the ideal MRI one; in the  absence of only gyroviscosity it is reduced to the ideal MTI one. As seen from the instability criterion (46), any dynamic instability appears to be dependent on  the signs of the angular velocity, temperature gradient and gyroviscous force which are the free energy sources. Departures from uniform rotation, and isothermality are indeed sources of dynamic instability. Gyroviscous force is coupled with angular velocity gradient. If  one refers to the definition of $\tilde{V}_{gyro}$, one clearly sees that gyroviscous force stabilizes or destabilizes the modes depending on the sign of   $\mathbf{\Omega}\cdot\mathbf{B}$. 

We consider the case of  astrophysical interest,  $d\Omega^2/d\ln R<0$. If $\mathbf{\Omega}$  and $\mathbf{B}_{z}$ are aligned in the same direction, i.e. $\mathbf{\Omega}\cdot\mathbf{B}>0$, the torque term which is the first factor of inequality (46) is  positive. The instability condition is determined by the sign of the radial force  term (second factor) in the inequality (46), that is, if the second factor is negative then instability arises:

\begin{equation}
k_{z}^2 v_{A}^{2}-\tilde{V}_{gyro}{d\Omega^2\over d\ln R}-{k_{z}^2\over k^2}{1\over\rho}{\partial P\over\partial R}{\partial \ln T\over\partial R}<-{k_{z}^2\over k^2}{d\Omega^2\over d\ln R}.
\end{equation}

This condition differs from the ideal MRI only by additional terms, i.e., gyroviscous and temperature gradient on the left hand side of the inequality (47). To ideal MRI, as long as ${d\Omega^2 / d\ln R}<0$ there  will  be an instability  for the small enough $k$ (Balbus \& Hawley 1998).

The gyroviscous force acts in the same direction with the magnetic tension force because of $\mathbf{\Omega}\cdot\mathbf{B}>0$. Therefore, the instability is suppressed because  the currents produced by  $\mathbf{E}\times\mathbf{B}$  drift  are being out of phase with the current of MRI eigenmode.
In the astrophysical situation ${\partial P/\partial R}<0$, because of the hydrostatic equilibrium.  If  one assumes that the temperature decreases in the direction of gravity, i.e. ${\partial\ln T/\partial R}>0$, then the dilute plasma is completely stable. If  one assumes that the temperature increases in the direction of gravity, i.e., ${\partial\ln T/\partial R}<0$, then the dilute plasma  may be unstable only if  the temperature gradient is very steep.

But in the situation where  $\mathbf{\Omega}$  and $\mathbf{B}_{z}$ are counter aligned, i.e., $\mathbf{\Omega}\cdot\mathbf{B}<0$,  the gyroviscous force acts in the opposite  direction to the magnetic tension force and enhances instability. Thus one may find unstable modes even if  ${\partial\ln T/\partial R}>0$,  this is because  the currents produced by $\mathbf{E}\times\mathbf{B}$   drift  are being in phase with the current of  MRI eigenmode. If  one assumes that the temperature increases in the direction of gravity, i.e., ${\partial\ln T/\partial R}<0$,  then MTI arises. In this case,  GvMRI and MTI have equal weights for dominance. It is expected that, especially at higher $k$ values, former is the dominant one. In many astrophysical plasmas, like cooling white dwarfs and neutron stars, hot accretion flows on compact objects release the gravitational potential energy and causes ${\partial\ln T/\partial R}<0$. But in the  plasmas having the temperature profile which there are no unstable mode to ideal MTI, i. e., ${\partial\ln T/\partial R}>0$, for example in the cooling flow clusters, GvMRI may operate.

By referring to the general instability criterion (45), one can argue that perturbations with $k_{R}$ always stabilizes because of the last term in the inequality (45). Interpretation of the criterion becomes difficult if we take into account the pitch angles different from $90^{\circ}$. In that case, the signs and the ratios of the different terms come into play. Therefore, it is more instructive to look at the numerical solutions of the dispersion relation.

\section{Numerical Solutions of the Dimensionless Dispersion Relation}

The dimensionless  dispersion relation is obtained that all the terms of Equation (39) is divided by $\Omega^5$:

\begin{equation}
{\gamma}^{5}+{b}_{4}{\gamma}^{4}+{b}_{3}{\gamma}^{3}+{b}_{2}{\gamma}^{2}+{b}_{1}{\gamma}+{b}_{0}=0,
\end{equation}

where \begin{equation}
{b}_4=6\tilde{V}_{par}^{n}Xs^2{k_{\perp}^2\over k^2}+\tilde{V}_{ther}^{n}Xs^2,
\end{equation}

\begin{eqnarray}
{b}_3=\tilde{V}_{gyro}^{n}X{d\ln\Omega^2\over d\ln R}(G+H)+4{k_{z}^2\over k^2}\tilde{V}_{gyro}^{n}X(2sy^2-A)
+2X+{k_{z}^2\over k^2}\tilde{\kappa}^2\nonumber\\
+{k_{z}^2\over k^2}(\tilde{V}_{gyro}^{n})^{2}X^2(2sy^2-A)^2
+\tilde{N}^{2}{k_{z}^2\over k^2}+6\tilde{V}_{par}^{n}\tilde{V}_{ther}^{n}X^2s^4{k_{\perp}^2\over k^2},
\end{eqnarray}

\begin{eqnarray}
{b}_2=6\tilde{V}_{par}^{n}s^2k_{z}^2X^2{k_{\perp}^2\over k^2}+\tilde{V}_{par}^{n}\tilde{V}_{gyro}^{n}X^2 3s^2G{d\ln\Omega^2\over d\ln R}{k_{\perp}^2\over k^2}
+{k_{z}^2\over k^2}\tilde{V}_{par}^{n}X{d\ln\Omega^2\over d\ln R}2cD\nonumber\\
+{k_{z}^2\over k^2}\tilde{V}_{gyro}^{n}6yX D{d\ln\Omega^2\over d\ln R}+(\tilde{V}_{gyro}^{n})^2X^2 3y D(2sy^2-A){d\ln\Omega^2\over d\ln R}{k_{z}^2\over k^2}\nonumber\\
+{k_{z}^2\over k^2}\tilde{V}_{par}^{n}X\tilde{N}^{2}2cD-{k_{z}^2\over k^2}\tilde{V}_{ther}^{n}Xs^{2}P_{M}T\nonumber\\
+\tilde{V}_{ther}^{n}Xs^{2}\left[\begin{array}{c}
2X+{k_{z}^2\over k^2}{\tilde{\kappa}}^2+{k_{z}^2\over k^2}(\tilde{V}_{gyro}^{n})^{2}X^2(2sy^2-A)^2+\tilde{V}_{gyro}^{n}X{d\ln\Omega^2\over d\ln R}(G+H)\\
+4{k_{z}^2\over k^2}\tilde{V}_{gyro}^{n}(2sy^2-A)\end{array}
\right],
\end{eqnarray}

\begin{eqnarray}
{b}_1=\left[X\left(1+\tilde{V}_{gyro}^{n}{d\ln\Omega^2\over d\ln R}G/2\right)+{k_{z}^2\over k^2}{d\ln\Omega^2\over d\ln R}\right]\left[X\left(1+\tilde{V}_{gyro}^{n}{d\ln\Omega^2\over d\ln R}(G/2+H)\right)\right]\nonumber\\
+{k_{z}^2\over k^2}y^2(\tilde{V}_{gyro}^{n})^2X^2 2D^2({d\ln\Omega^2\over d\ln R})^2\nonumber\\
+{k_{z}^2\over k^2}\tilde{N}^{2}\left[X+\tilde{V}_{gyro}^{n}{d\ln\Omega^2\over d\ln R}\left({G\over 2}+H\right)\right]-{k_{z}^2\over k^2}\tilde{V}_{ther}^{n}s^{2}\tilde{V}_{par}^{n}2cDP_{M}T\nonumber\\
+\tilde{V}_{ther}^{n}s^{2}\left[\begin{array}{c}
6\tilde{V}_{par}^{n}s^2X {k_{\perp}^2\over k^2}+\tilde{V}_{par}^{n}\tilde{V}_{gyro}^{n}3s^2G{d\ln\Omega^2\over d\ln R}{k_{\perp}^2\over k^2}
+{k_{z}^2\over k^2}\tilde{V}_{par}^{n}{d\ln\Omega^2\over d\ln R}2cD\\
+{k_{z}^2\over k^2}\tilde{V}_{gyro}^{n}6yD{d\ln\Omega^2\over d\ln R}
+(\tilde{V}_{gyro}^{n})^2 3yD(2sy^2-A){d\ln\Omega^2\over d\ln R}{k_{z}^2\over k^2}
\end{array}\right]
\end{eqnarray}

\begin{eqnarray}
{b}_0=\tilde{V}_{ther}^{n}Xs^{2}\left[\begin{array}{c}
\left[X+\tilde{V}_{gyro}^{n}{d\ln\Omega^2\over d\ln R}(G/2+H)\right]\left[X+\tilde{V}_{gyro}^{n}{d\ln\Omega^2\over d\ln R}{G\over 2}+{k_{z}^2\over k^2}{d\ln\Omega^2\over d\ln R}-{k_{z}^2\over k^2}P_{M}T\right]\\
+{k_{z}^2\over k^2}y^2(\tilde{V}_{gyro}^{n})^2 2D^2({d\ln\Omega^2\over d\ln R})^2
\end{array}
\right],
\end{eqnarray}

where $\gamma=\omega/\Omega$, $X=k_{z}^2v_{A}^{2}/\Omega^2$, $\tilde{V}_{gyro}^{n}=\tilde{V}_{gyro}/X$, $\tilde{V}_{par}^{n}={V}_{par}/\Omega X$, $\tilde{V}_{ther}^{n}={V}_{ther}/\Omega X$, $\tilde{N}=N/\Omega$,  $\tilde{\kappa}=\kappa/\Omega$, $P_{M}=(3/5M_{s}^2)(d\ln P/d\ln R)$ and $T=(d\ln T/d\ln R)$. $M_{s}=v_{\phi}/c_{s}$ is the Mach number.

Dilute plasma condition can be expressed as $\epsilon\equiv\omega_{ci}\nu_{i}\gg 1$. From this condition one can easily derives the inequality $\tilde{V}_{par}^{n}\gg \tilde{V}_{gyro}^{n}$.  Therefore, we assume $\tilde{V}_{par}^{n}=1000$. We consider the convectively stable plasmas,  i.e., $\tilde{N}^2>0$ and suppose $P_{M}=-1$.  Figure 4-8 show numerical solutions of the dimensionless dispersion relation (Equation 48) under the particular assumptions. All the figures are drawn as dimensionless growth rate versus wavenumber and  $\theta$ for the Keplerian rotational profile.

Figure 4 shows that there is an instability in small wavenumbers for  all pitch angle values when only  heat conduction is considered. Maximum growth rate of the instability is smaller than growth rate of ideal MRI ($0.75\Omega$). Indeed, this instability is a MTI which emerges in the presence of the helical magnetic geometry. Although the plasma with a temperature gradient increasing  outward is stable to the ideal MTI, it turns out to be unstable when threaded by a helical magnetic field.

\begin{figure}
\epsscale{.50}
\plotone{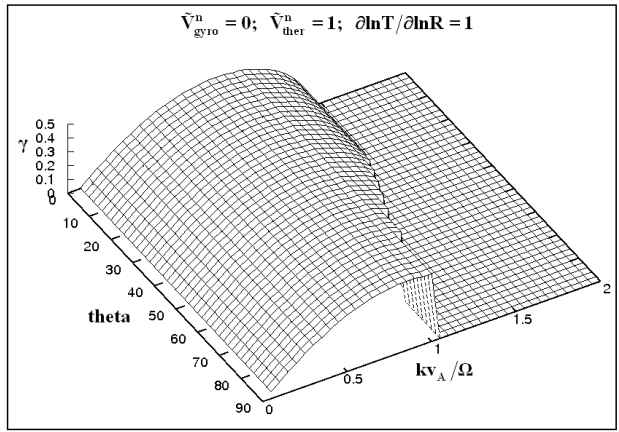}
\caption{The growth rates of MTI for helical magnetic geometry. This figure is drawn for $k_{R}=0$. Instability occurs only at small wavenumbers for all the pitch angles.   \label{fig1}}
\end{figure}

\begin{figure}
\epsscale{.90}
\plotone{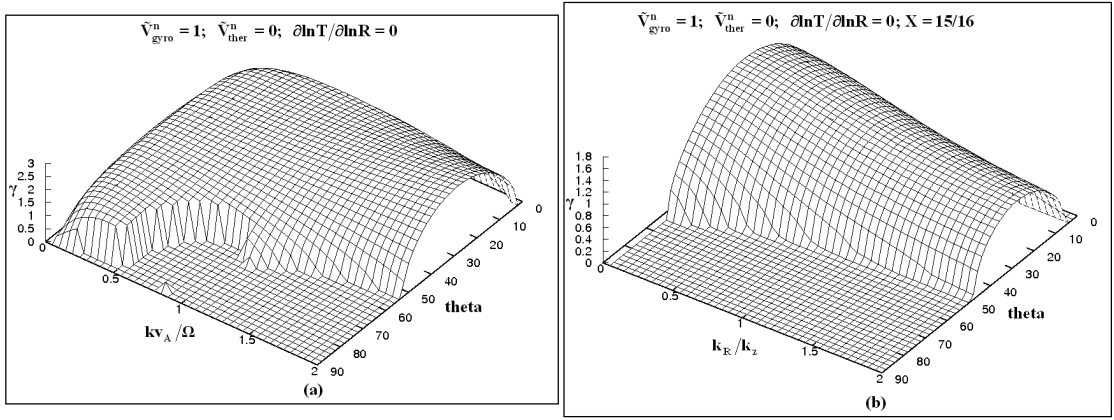}
\caption{The growth rates of GvMRI. Figure 5a is drawn for $k_{R}=0$. Figure 5b shows growth rates versus pitch angle and $k_{R}/k_{z}$ for  $X=(15/16)^{1/2}$. The mode with any  wavenumber is unstable for the pitch angles $\theta<50^{\circ}$.  \label{fig2}}
\end{figure}

\begin{figure}
\epsscale{.90}
\plotone{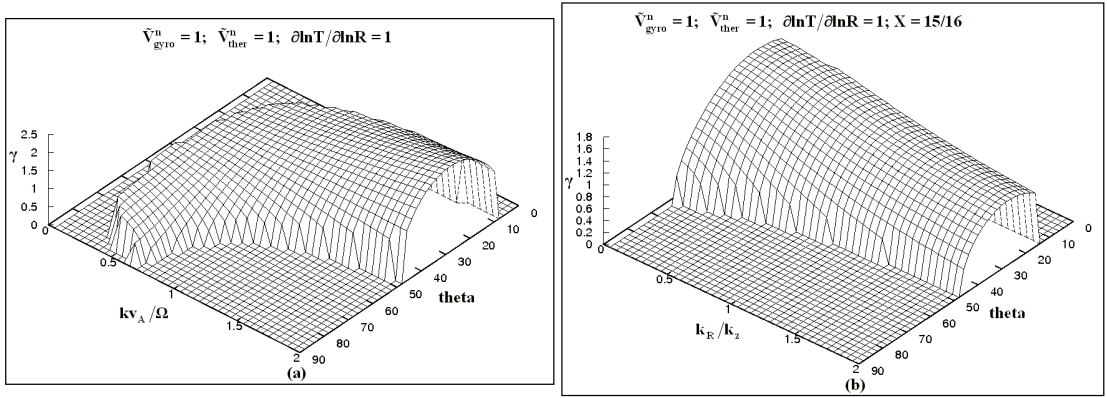}
\caption{The growth rates of the instability in the presence of  anisotropic transport effects. Figure 6a is drawn for $k_{R}=0$. Mode with large wavenumbers is unstable for the pitch angles  $10^{\circ}<\theta<50^{\circ}$. Figure 6b shows growth rates versus pitch angle and $k_{R}/k_{z}$ for  $X=(15/16)^{1/2}$. The radial wavenumber reduces the growth rates.\label{fig3}}
\end{figure}

\begin{figure}
\epsscale{.50}
\plotone{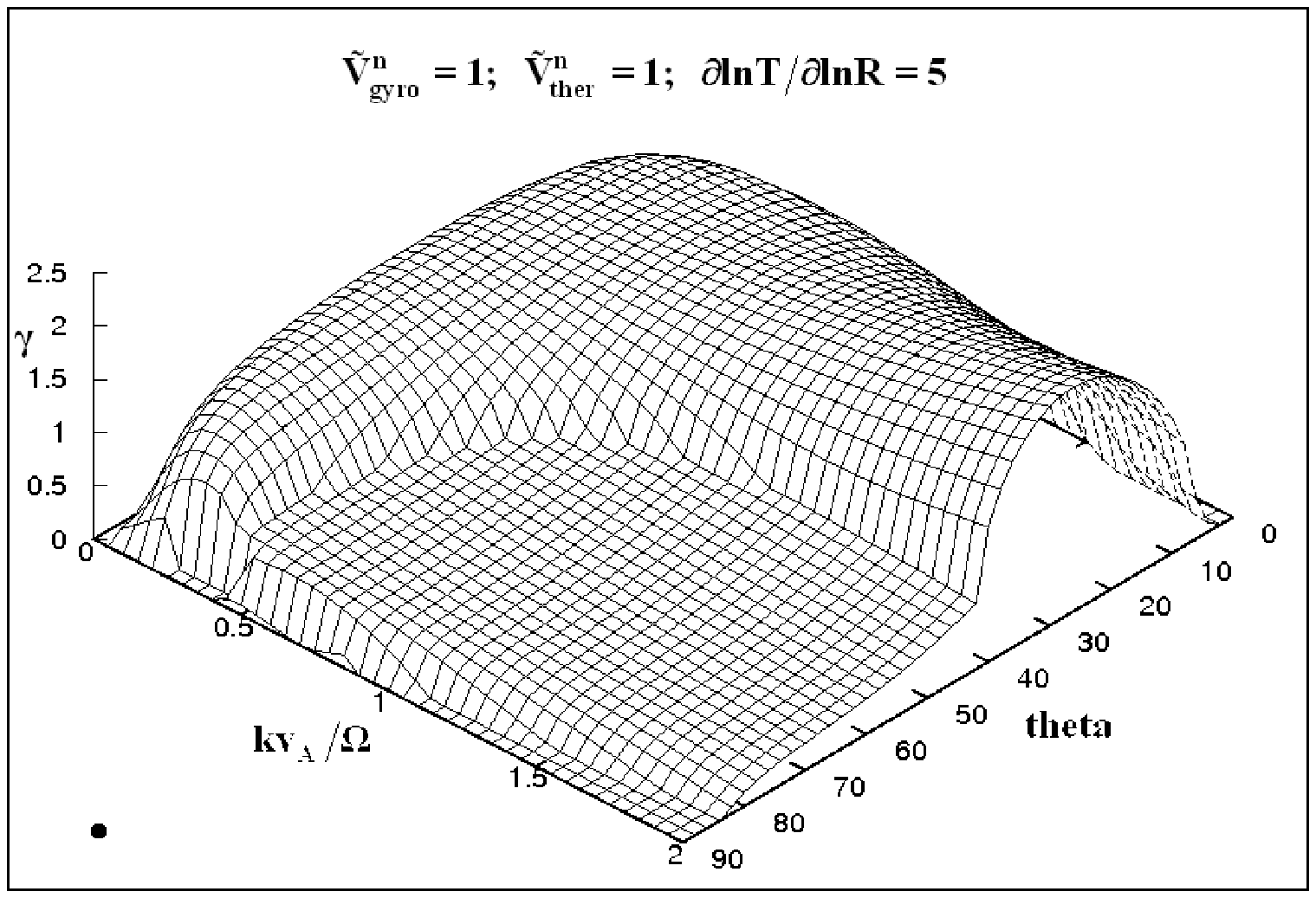}
\caption{The growth rates of the instability in the presence of the steep temperature gradient. Instability with higher growth rates arises for $\theta<40^{\circ}$. Besides, there is another but rather weak unstable region revealing itself as a mild crest for $\theta>40^{\circ}$.\label{fig4}}
\end{figure}

When the angular velocity vector and the $B_{z}$ component of the magnetic field are parallel ($\mathbf{\Omega}\uparrow\uparrow\mathbf{B}_{z}$),  gyroviscosity assumes positive values. Figure 5a is drawn only for the  $\tilde{V}_{gyro}^{n}=1$ and $k_{R}=0$ for the  case without heat conduction. This instability is a gyroviscous instability which mentioned in Paper I. The mode with small  wavenumber $<0.5$ is unstable for all the values of pitch angles. But, for the all wavenumbers, the unstable mode emerges only when  $\theta<50^{\circ}$. Maximum growth rate of instability is about $3\Omega$. Figure 4 and 5a  clearly show that GvMRI is more powerful and greater an instability than  magnetothermal one.

Figure 5b demonstrates  dimensionless growth rates versus pitch angle and $k_{R}/k_{z}$. In this figure, normalized wavenumber of the fastest growing mode of MRI is adopted as $X=(15/16)^{1/2}$. For the pitch angles $\theta<50^{\circ}$, the mode with any  wavenumber is unstable, but even if maximum growth rate is smaller than the case when $k_{R}=0$, it is still  greater than its correspondent in the  ideal MRI case.

The combined effects of the gyroviscosity and the heat conduction on the instability are seen in the Figure 6a. The cases of $\mathbf{\Omega}\uparrow\uparrow\mathbf{B}_{z}$ and $(d\ln T/d\ln R)>0$ are considered together. While the mode with  small wavenumbers  ($\sim<0.5$) for all the pitch angle values is stable, the one with  large wavenumbers becomes unstable only when the  pitch angle is within the range  of  $10^{\circ}<\theta<50^{\circ}$. Comparison of Figures 5a and 6a shows that the temperature gradient acts to reduce the growth rate of the instability. Also, it supresses the instability which  appears at small wavenumbers ($\sim<0.5$) and small pitch angles ($\theta<10^{\circ}$).

Figure 6b shows the growth rates versus pitch angle and $k_{R}/k_{z}$ for $X=(15/16)^{1/2}$.  The radial wavenumber of the unstable mode reduces the growth rates.

\begin{figure}
\epsscale{.90}
\plotone{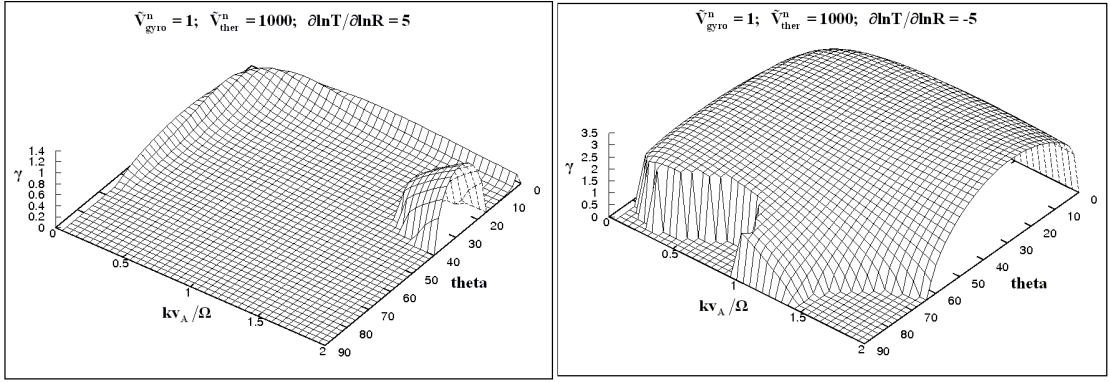}
\caption{The growth rates of the instability for the situation of $\omega_{cond}\gg\omega_{dyn}\gg\mathbf{k}\cdot\mathbf{v}_{A}$. Figure 8a and 8b are drawn for $k_{R}=0$. The instability is supressed when $T=5$ but when $T=-5$ almost at all the wavenumbers and pitch angles instability arises. \label{fig5}}
\end{figure}

When the  angular velocity vector and the $B_{z}$ component of the magnetic field are antiparallel ($\mathbf{\Omega}\uparrow\downarrow\mathbf{B}_{z}$),  gyroviscosity assumes negative values. For the situation $\tilde{V}_{gyro}^{n}<0$, an interesting situation arises. Instability sets in for the larger pitch angles $(\theta=60^{\circ}-90^{\circ})$. This figure is not included here for spatial economy, because it looks like mirror image of the one with $\mathbf{\Omega}\uparrow\uparrow\mathbf{B}_{z}$ (see Fig. 6a), but
the growth rates of instability are larger, i.e. $\sim 3\Omega$.  When we increase the value of $\tilde{V}_{gyro}^{n}$, irrespective of its sign, the maximum growth rates of the instability become larger, for example, for the $\tilde{V}_{gyro}^{n}=5$, the maximum growth rate is $\sim 6\Omega$.

Now, let us investigate the situation wherein the temperature gradient decreases outward. Instability behaves in the same manner, that is, the appearance of figures are the same but the maximum growth rate is comparatively smaller. 

If temperature gradient assumes a value of  $5$, and $\tilde{V}_{gyro}^{n}>0$  then instability arises for $\theta<40^{\circ}$. Besides, there is another but rather weak unstable region revealing itself as a mild crest for $\theta>40^{\circ}$. For $kv_A/\Omega = 2$, the growth rate is $0.45$ at the pitch angle $50^{\circ}$ and $0.13$ at $80^{\circ}$. The growth rates have bigger values at smaller wavenumbers (see Figure 7). If temperature gradient assumes a value of  $-5$, instability is seen for $\theta<40^{\circ}$, but there is no mild crest for $\theta>40^{\circ}$. Growth rates remains as the same. When $\tilde{V}_{gyro}^{n}<0$  and $\partial\ln T/\partial\ln R=5$, there is instability with the same growth rates at pitch angles $\theta>60^{\circ}$. For  $\theta<60^{\circ}$ again a mild crest appears as a second unstable region.

In the weak magnetic field limit,  $\omega_{cond}\gg\omega_{dyn}\gg\mathbf{k}\cdot\mathbf{v}_{A}$ ordering holds true. Here $\omega_{dyn}\sim(g/H)^{1/2}$ is the local dynamical frequency (Quataert 2008). If $\tilde{V}_{ther}^{n}=V_{ther}/X\Omega=(2k_{z}^2\chi T/5P)/(kv_{A})=\omega_{cond}/kv_{A}=1000$ and $\tilde{V}_{gyro}^{n}=1$, $\partial\ln T/\partial\ln R=1$, then the figure of the instability looks exactly like the Figure 5a, that is, only  the mode with wavenumbers smaller than $0.5$  is  unstable for all the possible pitch angles, and can not alter the maximum growth rate (see Fig. 5a). If this situation is compared with the Figure 6a which is assumed $\tilde{V}_{ther}^{n}=1$ we see that only the mode with very small wavenumbers is unstable for very large characteristic conduction frequency.

Steep temperature gradient and very large characteristic frequency about conduction assumption reveals a very interesting situation. While  instability is supressed for  $T=5$,  when $T=-5$,  it arises almost at all the wavenumbers and pitch angles (see Figs. 8a and 8b).

\subsection{Summary}

We have deduced the effects of various parameters on  the behavior of instability from numerical solutions of  the dimensionless  dispersion relation. These are 
 
\begin{enumerate}
 \item Maximum growth rate of the GvMRI ($2.5-3\Omega$) is greater than the growth rates of ideal MRI and MTI.

 \item The growth rate of unstable mode depends sensitively on the pitch angle and the gyroviscosity parameter $V_{gyro}/\Omega X=\tilde{V}_{gyro}^{n}$.

 \item   Gyroviscosity parameter is positive or  negative if  angular velocity vector and $B_{z}$ component of the magnetic field are oriented in the same or opposite sense. For these situations 
instability regions  in the figures are seen  as the mirror images of each other. In other words, the instability sets in at the smaller pitch angles ($\theta<50^{\circ}$) for the positive gyroviscosity parameter, but when this parameter is negative  instability sets in at the larger pitch angles ($50^{\circ}<\theta<90^{\circ}$).

 \item To ideal MTI, plasma is unstable in which the temperature increases in the direction of gravity, i.e. $\partial\ln T/\partial\ln R<0$. Whereas, we showed that there is at least one unstable mode for a plasma with $\partial\ln T/\partial\ln R<0$ or $\partial\ln T/\partial\ln R>0$ in the presence of  helical magnetic field. Instability occurs due to gyroviscous force.

 \item   What is the effect of temperature gradient  on the GvMRI? First of all, the maximum growth rate is reduced. Second, the mode with the very small wavenumbers ($k<0.5$) remain formally stable (see Fig 5a and 6a).

 \item Only in the presence of the steep temperature gradient, instability occurs at almost all the wavenumbers and pitch angles (see Fig. 7).

 \item In the presence of the steep temperature gradient and very large characteristic frequency about conduction, the temperature gradient term stabilizes or destabilizes depending on  whether it is positive  or negative (see Fig. 8). 

\end{enumerate}

\section{DISCUSSION AND CONCLUSION}

In a dilute plasma, ion cyclotron frequency  greatly exceeds ion-ion collision frequency and also electron mean free path is much larger than the gyroradius. Therefore, transport of
momentum and heat  by viscosity and thermal conduction is highly anisotropic with respect to the magnetic field orientation. In this regime, for a more accurate plasma model  the anisotropic transport terms, i.e. parallel viscosity, gyroviscosity and thermal conduction, must be taken into consideration in the MHD equations.  Many astrophysical plasmas display characteristics of the dilute plasma. For example, physical parameters and the conditions of the intracluster medium (ICM) of the galaxy clusters are being revealed by telescopes with high resolving powers.  Chandra X-ray Observatory measured X-ray luminosity ($10^{43}-10^{46}$ erg/s) emitted by the hot plasma in the ICM and based on this measurement the density distribution of ICM as a function of radius is determined (Parrish, Stone and Lemaster 2008). Peterson and Fabian (2006) reports that the typical densities are in the range of $10^{-3}-10^{-2}$  $cm^{-3}$ ; temperatures are $1-15$ keV. Carilli and Taylor (2002) estimated the magnetic field strength in the center of  ICM about $1-19$ $\mu G$ and $0.1-1.0$  $\mu G$ at the radius of 1 Mpc. With the above quoted values plasma beta is $\beta=8\pi P/B^{2}\sim200-2000$. This value implies that the ICM plasma is dilute and the mean free path of electrons is much longer than their gyroradius (Narayan \& Medvedev 2001).

Under these physical conditions, dilute and hot plasma in a differentially rotating disc is open to a blend of instabilities like, MRI, MTI and GvMRI. We proceed the analysis from simple to complex.  Let us start with a useful mechanical model  developed by Balbus \& Hawley (1992) for the MRI. This model consists two fluid elements which are tethered each other with vertical magnetic field. They are also embedded in a radial angular velocity gradient, so that the element orbiting at smaller radius rotates rapidly than the other element orbiting at larger radius. We suppose that these elements are residing at different vertical locations, but at the same radial locations initially. When these elements are radially displaced, the magnetic field will force them to return at their original locations. The outward element acquires angular momentum because it has a smaller velocity in its new radial location. But the inward element loses angular momentum, because it has a greater velocity in its new radial location. When the field lines become more stretched, since the inner element  continues to lose its angular momentum, it will fall farther inward; the other element moves farther out and gains the higher angular momentum. Thus,  process runs away and instability occurs. This is a classical MRI picture (Balbus \& Hawley 1992, 1998). The ions (because ion viscosity is higher than that of electrons in the dilute plasma)  in the fluid elements gyrate around the magnetic field lines and   will be under the influence of  a spatially varying electric field which arises from FLR effects. In the shear flow there is no rest frame. Therefore the elements in different locations are exposed to different electric fields ($\mathbf{E}=-\mathbf{v}\times\mathbf{B}/c=-R\Omega(R)B_{0}\sin\theta\mathbf{\hat R}$). Along a gyro-orbit, the length scale of electric field  is comparable to length scale of velocity gradient (Williams \& Jokipii 1991).  If the initially positions of elements is selected as a rest frame, the fluid velocity increases towards smaller $R$ values. So, the inward element will see larger electric field because it has greater velocity  according to the outward element. Since the drift is controlled by the magnitude of the electric field, the fluid element with a larger relative velocity will drift more rapidly. Because the magnetic field which tethers fluid elements to each other acts as spring-like force, the rapidly increasing element separation gives rise to growing spring tension.  Thus, process runs away  and instability occurs quickly. This is a  GvMRI picture. 
   
The modified Hill equations by the inclusion of gyroviscosity, thermal conduction and parallel viscosity can help one to get a better physical understanding of instability. In the absence of all three dynamical effects, one recovers the original set of equations describing the MRI (Balbus \& Hawley 1992, 1998). In equations (54) and (55) below, $\xi_{R}$ and $\xi_{\phi}$ are the radial and azimuthal displacements of fluid elements.

\begin{eqnarray}
{\partial^2 \xi_{R}\over\partial t^2}-2\Omega\left(1-{1\over2}\tilde{V}_{gyro}A\right){\partial \xi_{\phi}\over\partial t}\nonumber\\
=-\left[\begin{array}{c}
\left(k_z v_{A}\right)^2 +{d\Omega^2\over d\ln R}\left(1+\tilde{V}_{gyro}{G\over 2}\right)\\
+\left(N^2\omega-\Omega^2{V}_{ther}\sin^2\theta P_{M}T\right)\left(\omega+{V}_{ther}\sin^2\theta\right)^{-1}
\end{array}
\right]\xi_{R}
\end{eqnarray}

\begin{eqnarray}
{\partial^2 \xi_{\phi}\over\partial t^2}+2\Omega\left(1-{1\over2}\tilde{V}_{gyro}A\right){\partial \xi_{R}\over\partial t}=-\left[\begin{array}{c}
\left(k_z v_{A}\right)^2 +\tilde{V}_{gyro}{d\Omega^2\over d\ln R}\left({G\over 2}+H\right)\\
+\left(2\omega {V}_{par}\cos\theta D \right)
\end{array}
\right]\xi_{\phi}
\end{eqnarray}

where  all the abbreviations are the same as above.

The right hand sides (r.h.s.) of the equations (54) and (55) represent the torque applied to the fluid elements which correspond to ``spring'' constants in radial and azimuthal directions. Comparison of r.h.s. of the equations (54) and (55) show that the gyroviscosity couples to the differential rotational, the thermal conduction couples to the Coriolis force, the radial gradients of the temperature and the pressure. These complex couplings make the roles of the anisotropic  forces intangible.  We solve equation (55) for $\xi_{\phi}$ and then substitute it into equation (54) to find the acceleration of the perturbed fluid element in the radial direction. In order the hot, dilute and differentially rotating disc to be unstable there should be a net outward acceleration:

\begin{eqnarray}
{\partial^2 \xi_{R}\over\partial t^2}=-\left(k_z v_{A}\right)^2\xi_{R}\nonumber\\
-\left[\kappa^2+\left(N^2\omega-\Omega^2 {V}_{ther}\sin^2\theta P_{M}\right){T\over{\omega+{V}_{ther}\sin^2\theta}}\right]\xi_{R}\nonumber\\
-\left[{d\Omega^2\over d\ln R}\tilde{V}_{gyro}{G\over 2}+4\Omega^2\left(-\tilde{V}_{gyro} A+{1\over 4}\tilde{V}_{gyro}^{2}A^2\right)-8\Omega^2\left(1-{1\over 2}\tilde{V}_{gyro}A\right)^2{Y\over (\omega^2+Y)}\right]
\xi_{R}
\end{eqnarray}

where $Y=\left[\left(k_z v_A\right)^2+\tilde{V}_{gyro}{d\Omega^2\over d\ln R}({G\over 2}+H)+2D\omega {V}_{par}\cos\theta \right]$.

Accelaration given by the equation (56) can be written as a sum of three terms respectively, \textit{i.e.}, $\partial^2\xi_R /\partial t^2={a}_{T}={a}_M+{a}_{H,th}+{a}_{M,gv}$ . 

${a}_M$, given by the equation (56) is the acceleration term arising from the magnetic tension force which is always stabilizing. The second acceleration term ${a}_{H,th}$ is due to the radial bouyancy force and  anisotropic thermal conduction.  And it is a pure hydrodynamic term. This term is related to the convective instability  in which the source of free energy  is the temperature gradient. In the absence of the viscous force, when the temperature increases in the direction of gravity, this term has a destabilizing effect and when the temperature decreases in the direction of gravity, it has a stabilizing effect. The third acceleration term ${a}_{M,gv}$ is due to the radial gyroviscous force. So, it is related to the GvMRI. This term  depending on the pitch angle, the wavenumbers, gyroviscous force and the parallel viscosity may be either positive or negative sign which corresponds to the destabilizing and stabilizing effect, respectively.

\begin{figure}
\epsscale{.65}
\plotone{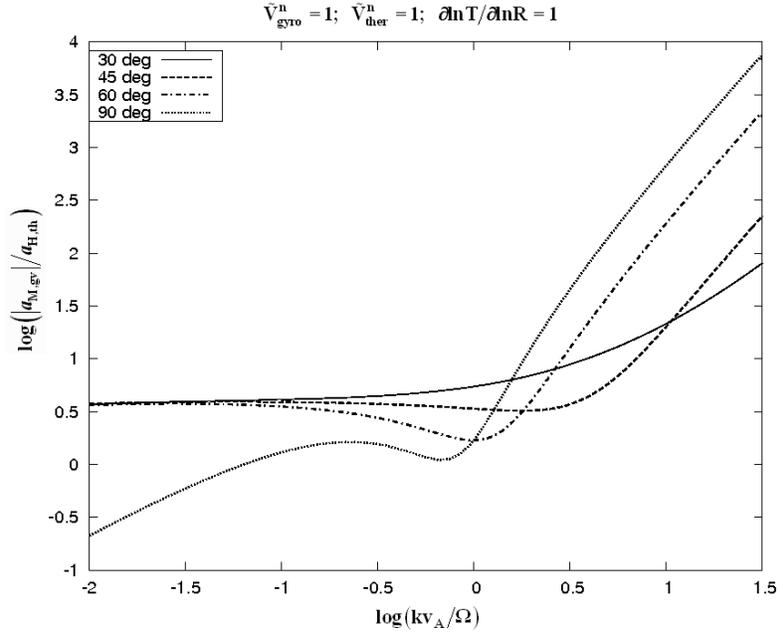}
\caption{Figure is drawn as the logaritmic normalized wavenumber versus logaritmic  ratio of accelerations $log(a_{M,gv}/a_{H,th})$ for convectively stable plasma, i.e. $N^2=1$,  in the presence of the three dynamical forces, i.e., parallel viscosity, thermal conduction and gyroviscosity. Throughout the whole wavelength range, acceleration term ${a}_{M,gv}$ which due to  gyroviscous force is dominant. Also, this acceleration increases rapidly in the large wavenumbers.  \label{fig6}}
\end{figure}

The relative importance of the two terms which can be stabilizing or destabilizing  are depicted in Figure 9. Figure is drawn as the logaritmic  normalized wavenumber versus logaritmic ratio of  accelerations $log(a_{M,gv}/a_{H,th})$ for convectively stable plasma, i.e., $N^2=1$.
If ${a}_{M,gv}\geq{a}_{H,th}$, the all  unstable modes are GvMRI modes. If ${a}_{M,gv}<{a}_{H,th}$, the unstable modes are called as thermal modes. 
As shown in the Figure 9, MTI by driven heat conduction  is dominant in the  very small wavenumbers only for the $\theta=90^{\circ}$, i.e., in considering only vertical magnetic field. In  all the possible situations throughout the whole wavelength range, acceleration term ${a}_{M,gv}$ due to  the gyroviscous force is dominant. Moreover, this acceleration  increases rapidly in the large wavenumbers. $({a}_{H,th})\xi_{R}$ term may be positive for the given values.  Still, this term does not make a contribution to instability even if $\tilde{V}_{ther}^{n}=1000$.

The above analysis shows that, regardless of the sign of the temperature gradient, a weakly magnetized and convectively stable dilute plasma harbouring the combined effects of gyroviscosity, parallel viscosity and thermal conduction  is unstable due to the gyroviscous force. Extension of the unstable regions and the growth rates of GvMRI depends sensitively on the pitch angle and the gyroviscosity parameter.
When the angular velocity vector and $B_{z}$ component of the magnetic field are parallel ($\mathbf{\Omega}\uparrow\uparrow\mathbf{B}_{z}$),  gyroviscosity assumes positive values. When the angular velocity vector and $B_{z}$ component of the magnetic field are antiparallel ($\mathbf{\Omega}\uparrow\downarrow\mathbf{B}_{z}$),  gyroviscosity assumes negative values. Former, while instability sets in for the smaller pitch angles 
$(\theta < 50^{\circ})$, later it sets in for the larger pitch angles $(\theta=60^{\circ}-90^{\circ})$. Maximum growth rates of the unstable mode is approximately $2.5\Omega$ and $3\Omega$ respectively. These values are higher than the ones of magnetorotational instability $(0.75\Omega)$ and the ones of magnetothermal instability ($0.5\Omega$, see Figure 4).
If the $k_{R}$ wavenumber is taken into consideration, the instability region becomes narrower and the growth rate of the instability is reduced.

Briefly, GvMRI is the dominant one in the blend of all three, that is pure MRI, MTI, gyroviscous modified MRI. Because the magnitude and the direction of the temperature gradient do not exert great effect on the unstable mode, this instability may work   at  all the astrophysical media including dilute plasma.

\acknowledgments
I thank E. Rennan Pek\"unl\"u for helpful discussions.
I am particularly grateful to the referee for  constructive comments which  led to a substantial improvement of this manuscript. This research was supported by the Scientific \& Technological Research Council of Turkey (T\"UB\.ITAK).




\appendix

\section{Two-fluid Equations}

The two-fluid equations together with the Faraday and Amp\'{e}re laws, respectively, are given by
(Braginskii 1965):

\begin{equation}
{d n_{s}\over dt}=-n_{s}\nabla\cdot\mathbf{v}_s,
\end{equation}

\begin{equation}
m_s n_s {d{\mathbf{v}_s}\over dt}=-\nabla P_s-\nabla\cdot\mathbf{\Pi}_s+q_s n_s \left(\mathbf{E}+{\mathbf{v}_s\times\mathbf{B}\over c}\right)+m_s n_s \mathbf{g},
\end{equation}

\begin{equation}
{3\over2}n_s {d T_s \over dt}=-P_s\left(\nabla\cdot\mathbf{v}_s\right)-\nabla\cdot\mathbf{Q}_s,
\end{equation}

\begin{equation}
{\partial{\mathbf{B}}\over\partial{t}}=-c\nabla\times\mathbf{E},
\end{equation}

\begin{equation}
\mathbf{J}={c \over 4\pi}\nabla\times\mathbf{B}=e \left(Zn_i \mathbf{v}_i-n_e \mathbf{v}_e \right),
\end{equation}

where the subscript ``s'' stands for electrons and ions,
$m$ is the mass, $n$ is the number density, $\mathbf{v}$ is the velocity of the  plasma components,  $P$ is the scalar pressure,   $\mathbf{\Pi}$  is the stress tensor, $q$ is the particle charge, $\mathbf{E}$ and $\mathbf{B}$ are the electric and magnetic fields, $\mathbf{g}$  is the gravitational acceleration, $\mathbf{Q}$ is the heat flux, $c$ is the speed of the light, $e$ is the electric charge ($q=e$), $Z$  is the charge state and   $d / dt=\partial/\partial t+\mathbf{v}\cdot\nabla$ is a Lagrangian derivative.

Most plasmas of interest are electrically neutral over sufficiently long distance and timescales. One assumes that the Debye length $\left(\lambda_D=\left(kT/4\pi ne^2\right)^{1/2}\right)$ is smaller than all the relevant spatial scales $\lambda_D\ll L$.                                            
Hence ion and electron densities are essentially equal, i.e. $n_e=Zn_i$,  quasi-neutrality. Then the mass
continuity equation is written only for ion number density:

\begin{equation}
{d n_{i}\over dt}=-n_{i}\nabla\cdot\mathbf{v}_i.
\end{equation}

Let us multiply the equation (A6) with $m_i$. Since $\rho=m_i n_i$, the equation (A6) now reads 

\begin{equation}
{d \rho\over dt}+\rho\nabla\cdot\mathbf{v}=0.
\end{equation}

Let us first write $\mathbf{E}$ by using from the electron momentum equation:

\begin{equation}
\mathbf{E}={1\over en_e}\left[-m_e n_e {d{\mathbf{v}_e}\over dt}-\nabla P_e-\nabla\cdot\mathbf{\Pi}_e+m_e n_e \mathbf{g} \right]-{\mathbf{v}_e\times\mathbf{B}\over c}.
\end{equation}

Let us eliminate $\mathbf{E}$ (Equation A8) by substituting from the electron momentum equation into ion momentum
equation:

\begin{eqnarray}
m_i n_i {d{\mathbf{v}_i}\over dt}=-\nabla P_i-\nabla\cdot\mathbf{\Pi}_i+Ze n_i {1\over en_e}\left[-m_e n_e {d{\mathbf{v}_e}\over dt}-\nabla P_e-\nabla\cdot\mathbf{\Pi}_e+m_e n_e \mathbf{g} \right] \nonumber \\
-Ze n_i {\mathbf{v}_e\times\mathbf{B}\over c}+Ze n_i{\mathbf{v}_i\times\mathbf{B}\over c}+m_i n_i \mathbf{g}.
\end{eqnarray}

Let us put $\rho=m_i n_i$, $P=P_e+P_i$, $\Pi=\Pi_e+\Pi_i$ and assume that $m_e/m_i\sim 0$, because $m_e\ll m_i$.
With the substitutions and using Amp\'{e}re laws, the ion momentum equation becomes

\begin{equation}
\rho{d{\mathbf{v}}\over dt}=-\nabla P-\nabla\cdot\mathbf{\Pi}+{\mathbf{J}\times\mathbf{B}\over c}+\rho\mathbf{g}.
\end{equation}

Let us substitute $\mathbf{E}$ (Equation A8) into the magnetic induction equation as given (A4):

\begin{equation}
{\partial{\mathbf{B}}\over\partial{t}}=-c\nabla\times \left[{cm_eZ\over e n_e m_i}\rho\left(\mathbf{g}-{d \mathbf{v}_e \over d t}\right)-{1\over en_e}\nabla P_e-{1\over en_e}\nabla \cdot \mathbf{\Pi}_e-\mathbf{v_e}\times\mathbf{B}\right].
\end{equation}

Since it is assumed that $n_e=Zn_i$ then $\mathbf{J}={(c/ 4\pi)}\nabla\times\mathbf{B}=e n_e\left(\mathbf{v}_i-\mathbf{v}_e \right)$ may be written in this form. From this equation one obtains $\mathbf{v}_e=\mathbf{v}_i-\mathbf{J}/en_e$.
After substituting $\mathbf{v}_e$ and using $m_e/m_i\sim 0$, one obtains magnetic induction equation as given
below:

\begin{equation}
{\partial{\mathbf{B}}\over\partial{t}}=\nabla\times \left(\mathbf{v_i}\times\mathbf{B}-{1\over en_e}\mathbf{J}\times \mathbf{B}+{c\over en_e}\nabla P_e+{c\over en_e}\nabla\cdot\mathbf{\Pi}_e\right).
\end{equation}

The second term on the right-hand side of the equation (A12) is Hall effect. This term is negligible
because in the present investigation
$\beta\gg 1$ limit  is considered ($\beta$ is the ratio of the gas pressure to the magnetic pressure). The third term on the right-hand side of the equation (A12) is thermodiffusion term.
The ions carry the most of the momentum due to their higher masses. Therefore, the third and the last term on the
right-hand side of the equation (A12) are negligible. So that resulting magnetic induction equation is
given by

\begin{equation}
{\partial{\mathbf{B}}\over\partial{t}}=\nabla\times \left(\mathbf{v}\times\mathbf{B}\right).
\end{equation}

If the ions and electrons are in thermal equilibrium, then $T_e\simeq T_i\simeq T$. Let us add ion energy equation and electron energy equation:

\begin{equation}
(n_e+n_i){dT\over dt}=(Z+1)n_i{dT\over dt}=-{2\over 3}\nabla\cdot\mathbf{v_i}(P_e+P_i)+{2\over 3}P_e\nabla\cdot{\mathbf{J}\over en_e} -{2\over 3}\nabla\cdot\mathbf{Q}.
\end{equation}

One assumes an ideal gas equation of state. Therefore one may write  $P=(n_e+n_i)T$. One obtains the resulting energy equation
after some algebraic operations:

\begin{equation}
{dP\over dt}+{5\over 3}P\left(\nabla\cdot\mathbf{v}\right)=-{2\over 3}\nabla\cdot\mathbf{Q}.
\end{equation}

\section{The Negligible Effect of Stress Tensor on The Equilibrium State}

On the equilibrium state, the contribution from parallel viscosity components of stress tensor is

\begin{equation}
\nabla\cdot\mathbf{\Pi}_{0}^{v}=0.96{P_{i}\over 2\nu_{i}}\hat{R}{\partial\over\partial R}\cdot\left(\mathbf{I}-3\mathbf{\hat{b}}\mathbf{\hat{b}}\right)\left(\mathbf{\hat{b}}\cdot\mathbf{W}\cdot\mathbf{\hat{b}}\right).
\end{equation}

The rate of strain tensor has two components in the equilibrium state, $W_{R\phi}=W_{\phi R}=d\Omega/d\ln R$.
The unit vector along the magnetic field is $\mathbf{\hat{b}}=\hat{\phi} b_{\phi}+\hat{z} b_{z}$. For this case,

\begin{equation}
\left(\mathbf{\hat{b}}\cdot\mathbf{W}\cdot\mathbf{\hat{b}}\right)=\left(\hat{R} b_{\phi}W_{\phi R}\right)\cdot \left(\hat{\phi} b_{\phi}+\hat{z} b_{z}\right)=0.
\end{equation}

Accordingly, there is no  contribution of the parallel viscosity to the equilibrium state.

In the equilibrium state, the contribution from gyroviscosity components of stress tensor is 

\begin{equation}
\nabla\cdot\mathbf{\Pi}^{gv}_0={P_{i}\over 4\omega_{ci}}\hat{R}{\partial\over\partial R}\cdot\left[\mathbf{\hat{b}}\times\mathbf{W}\cdot\left(\mathbf{I}+3\mathbf{\hat{b}}\mathbf{\hat{b}}\right)+\left[\mathbf{\hat{b}}\times\mathbf{W}\cdot\left(\mathbf{I}+3\mathbf{\hat{b}}\mathbf{\hat{b}}\right)\right]^{T}\right].
\end{equation}

Since $\mathbf{\hat{b}}\times\mathbf{W}\cdot 3\mathbf{\hat{b}}\mathbf{\hat{b}}= 0$, the gyroviscosity component is given by 

\begin{equation}
\nabla\cdot\mathbf{\Pi}^{gv}_0={P_{i}\over 4\omega_{ci}}\hat{R}{\partial\over\partial R}\cdot\left(\mathbf{\hat{b}}\times\mathbf{W}\cdot\mathbf{I}+\left[\mathbf{\hat{b}}\times\mathbf{W}\cdot\mathbf{I}\right]^{T}\right).
\end{equation}

Using vector and dyadic relationships, one obtains 

\begin{equation}
\nabla\cdot\mathbf{\Pi}_{0}^{gv}=-{P_i\over2\omega_{ci}}sin\theta\left[{\partial\over\partial R}\left({d\Omega\over d\ln R}\right)+{2\over R}{d\Omega\over d\ln R}\right].
\end{equation}

The hydrostatic equilibrium  equation (30) should be written as 

\begin{equation}
{\nabla P_{0}\over\rho_{0}}=\mathbf{g}+R\Omega^2+{P_i\over2\omega_{ci}\rho}sin\theta\left[{\partial\over\partial R}\left({d\Omega\over d\ln R}\right)+{2\over R}{d\Omega\over d\ln R}\right].
\end{equation}

Now, Equation (B6) may be rewritten for the gyroviscosity parametre $\tilde{V}_{gyro}^{n}=\Omega P_{i}/4\omega_{ci}\rho v_A^{2}$:

\begin{equation}
{\nabla P_{0}\over\rho_{0}}=\mathbf{g}+R\Omega^2+\tilde{V}_{gyro}^{n} 2v_A^{2}sin\theta\left[{1\over \Omega}{\partial\over\partial R}\left({d\Omega\over d\ln R}\right)+{1\over \Omega}{2\over R}{d\Omega\over d\ln R}\right].
\end{equation}

The first term in the bracket in the equation (B7)  is

\begin{equation}
I={1\over \Omega}{\partial\over\partial R}\left({d\Omega\over d\ln R}\right)={2\Omega\over 2\Omega^2}{\partial\over\partial R}\left({d\Omega\over d\ln R}\right)={1\over 2\Omega^2}\left[{\partial\over\partial R}\left(2\Omega{d\Omega\over d\ln R}\right)-2{\partial\Omega\over\partial R}{d\Omega\over d\ln R}\right],
\end{equation}

because

\begin{equation}
{\partial\over\partial R}\left(2\Omega{d\Omega\over d\ln R}\right)=2\Omega{\partial\over\partial R}\left({d\Omega\over d\ln R}\right)+2{\partial\Omega\over\partial R}{d\Omega\over d\ln R}.
\end{equation}

After some algebraic manipulation Equation (B8) is 

\begin{equation}
I={1\over 2\Omega^2}\left[{\partial\over\partial R}\left(\Omega^2{d\ln\Omega^2\over d\ln R}\right)-\Omega{\partial\Omega\over\partial R}{d\ln\Omega^2\over d\ln R}\right],
\end{equation}

then,

\begin{equation}
I={1\over 2\Omega^2}\left[\Omega{\partial\Omega\over\partial R}{d\ln\Omega^2\over d\ln R}+\Omega^2{\partial\over\partial R}\left({d\ln\Omega^2\over d\ln R}\right)\right],
\end{equation}

then,

\begin{equation}
I=\left[{1\over 4R}{\partial\ln\Omega^2\over\partial\ln R}{d\ln\Omega^2\over d\ln R}+{\partial\over\partial R}\left({d\ln\Omega^2\over d\ln R}\right)\right].
\end{equation}

The second term in the bracket in the equation (B7)is

\begin{equation}
II={1\over \Omega}{2\over R}{d\Omega\over d\ln R}={\Omega\over \Omega^2}{2\over R}{d\Omega\over d\ln R}={1\over R}{d\ln\Omega^2\over d\ln R}.
\end{equation}

Now, substituting the Equations (B12) and (B13) into the equilibrium state equation, we find

\begin{equation}
{\nabla P_{0}\over\rho_{0}}=\mathbf{g}+R\Omega^2+\tilde{V}_{gyro}^{n} 2v_A^{2}sin\theta\left[{1\over 4R}{\partial\ln\Omega^2\over\partial\ln R}{d\ln\Omega^2\over d\ln R}+{\partial\over\partial R}\left({d\ln\Omega^2\over d\ln R}\right)+{1\over R}{d\ln\Omega^2\over d\ln R}\right].
\end{equation}

In a Keplerian disc, $\partial\ln\Omega^2/\partial\ln R=-3$. Thus, the equilibrium state equation is obtained as
 
\begin{equation}
{\nabla P_{0}\over\rho_{0}}=\mathbf{g}+R\Omega^2-{6\over 4}\tilde{V}_{gyro}^{n} {v_A^{2}\over R}sin\theta.
\end{equation} 

One may consider $v_A^{2}/R\propto B^2/R$. In a dilute plasma, especially in the intracluster medium of galaxy clusters (see Chapter 4), the magnetic field is extremely weak and $R$ is relatively very large. Therefore, the contribution of the stress tensor to the equilibrium state is negligibly small. And the equilibrium state is given by

\begin{equation}
{\nabla P_{0}\over\rho_{0}}=\mathbf{g}+R\Omega^2.
\end{equation}




\clearpage



\end{document}